\documentclass[twocolumn,showpacs,preprintnumbers,amsmath,amssymb,nofootinbib]{revtex4}
\usepackage{graphicx,color}
\usepackage{bm} 
\newcommand{\qq}{{\bf q}}

\newcommand{\rr}{{\bf r}}

\newcommand{\be}{\begin{equation}}
\newcommand{\ee}{\end{equation}}
\newcommand{\ba}{\begin{eqnarray}}
\newcommand{\ea}{\end{eqnarray}}
\newcommand{\bse}{\begin{subequations}}
\newcommand{\ese}{\end{subequations}}
\newcommand{\beq}{\begin{eqnarray}}
\newcommand{\eeq}{\end{eqnarray}}
\newcommand{\ds}{\displaystyle}

\newcommand{\VMNM}{V_{M\!N\!M}}
\newcommand{\vMNM}{v_{M\!N\!M}}

\definecolor{darkb}{rgb}{0.,0.0,0.8}

\begin{document}
\title{An equation of state for expanded metals}
\author{W. Schirmacher$^1$, W.-C. Pilgrim$^2$ and F. Hensel$^2$} 
 
\affiliation{%
	\mbox{$^1$%
Institut f\"ur Physik, Johannes-Gutenberg-Universit\"at Mainz, Staudinger Weg 9,
D-55099 Mainz, Germany;}
$^2$%
Fachbereich 15, Chemie, Physikalische Chemie, 
Fachbereich Chemie, 
Philipps-Universit\"at Marburg, Hans-Meerwein-Strasse 4
D-35032 Marburg, Germany.
}
\begin{abstract}
We present a  model equation of states for expanded metals, which contains
a pressure term due to a screened-Coulomb potential with
a screening parameter reflecting the Mott-Anderson metal-to-nonmetal transition.
As anticipated almost 80 years ago by Zel'dovich and Landau, this term
gives rise to a second coexistence line in the phase diagram,
	indicating a phase separation between a metallic and a nonmetallic liquid.
\end{abstract}
\maketitle
\section{Introduction}
For almost 80 years
the relation of the metal-nonmetal and liquid-vapor transition of
expanded metals is not understood 
\cite{landau43,yonezawa82,hensel89,henselbook,hensel10}, despite
several efforts in the last decades \cite{henselbook,bobrov13}.

In their original paper \cite{landau43}
Zel'dovich and Landau (ZL) present arguments that in expanded metals, in particular
mercury, in addition to the usual liquid vapor coexistence line,
a second coexistence line (with a second critical point)
exists in the $(p,T)$ 
(pressure-temperature)
phase diagram,
which involves a phase separation between a metallic and nonmetallic liquid.

Experiments in expanded mercury \cite{henselbook,hensel10,inui06,inui07,ruland09} and, more recently
in expanded rubidium \cite{pilgrim18} revealed a density regime,
in which there is evidence for an emulsion of a metallic and a
nonmetallic phase,
thus confirming the ideas of ZL.

The arguments of ZL had been based on the assumption that
at zero temperature a discontinuous metal-nonmetal transition takes
place, which continues to be present at elevated temperatures. 

At these times the only known
mechanism for a crossover from a metal to a dielectric
was the de-overlapping of bands. ZL argued that a
continuous band-de-overlapping transition cannot take place,
because in the insulating state an excited electron across the gap
interacts via the Coulomb interaction with the hole left behind and
thus enhances the gap, when, in the absence of the interaction
it would go continuously towards zero.

The role of the electonic Coulomb
and exchange interaction on the Metal-nonmetal transition has been addressed
extensively by Mott (Mott transition)
\cite{mott49,mottdavis71,mott82,mott90}. He realized that these combined
interactions produce two separate bands of electrons with opposite spins,
giving rise to antiferromagnetic ordering in the insulating state. This
scenario, which can be described by the Hubbard model \cite{hubbard63},
was called Mott transition. Mott believed that the metal-nonmetal transition
was discontinuous and postulated the existence of a minimal metallic
conductivity. These ideas have been 
quantified by \textcite{yonezawa82} for calculations of the thermodynamic
properties of expanded metals
based on the Hubbard model and the coherent-potential approximation.
In these calculations an unstable density regime due to
the metal-nonmetal transition was identified.

Anderson \cite{anderson58} showed that disorder can be another
reason for a metal-nonmetal transition. This was first
demonstrated for non-interacting electrons (Anderson transition).
It was then shown \cite{anderson79} that the Anderson transition
is an interference phenomenon and 
could be indentified as a second-order (i.e. continuous) phase
transition with a non-thermal control parameter, namely the
amount of spatial potential fluctuations, seen by an electron.
The Anderson scenario, i.e. an electron in a random potential,
could be mapped onto the nonlinear sigma model of
planar ferromagnets \cite{wegner76,schafer80,mckane81}, which obeys the
same scaling as the Anderson transition.
In this 
field theory the density of states at the Fermi level $\mu$ was
identified as the order parameter, but the critical exponent $\beta$
turned out to be zero, so that in the ``non-ordered state'', the
nonmetallic, $\mu$ remained finite. That the Anderson transition
is continuous was confirmed by 
experiments on doped semiconductors
\cite{rosenbaum80}.

The nonlinear-sigma-model description was generalized to include
the electronic Coulomb interaction (Mott-Anderson transition)
\cite{finkelstein83,belitz94,belitz95}.
In the presence of the interaction the critical order-parameter
exponent $\beta$ became non-zero, so that $\mu$ acquired its usual
order-parameter role. A severe drawback of the nonlinear-sigma-model
approach is that it is based on treating the variance of the potential
fluctuations as small parameter, so the theory is restricted
to the weak-disorder limit.

A quite different approach, which does not suffer from this shortcoming, 
is the dynamical mean-field theory (DMFT) based on the Hubbard model
\cite{metzner89,janis89,georges92} and turned out to be a reliable
means for treating correlated electronic systems and the Mott transition.

By including disorder into the Hubbard model it proved possible to
treat the Mott-Anderson transition by means of the  DFMT
\cite{dobro97,byczuk05,dobro10,byczuk10}.
These developments showed that the {\it local} single-site density of
states $\mu_i$ of the disordered interacting electron system exhibits
a very broad distribution. As a consequence
the arithmetic mean $\langle\mu\rangle$ and
the geometric one $\langle\mu\rangle_g$ was shown to become very different
in the limit of strong disorder. It was shown, that, in fact
$\langle\mu\rangle_g$ is critical at the Anderson-Mott transition,
even in the Anderson case, whereas $\langle\mu\rangle$ is not.
\cite{dobro97,byczuk05,dobro10,byczuk10}.

In the present contribution we shall show that such a continuous
Mott-Anderson transition of the electrons produces
in an expanded metal
an instability in the density regime near the transition, and thus a
second phase-separation line between a metallic and nonmetallic liquid
phase, as anticipated by ZL.

We argue that in expanded metals
the transition scenario for the electrons is a qualitatively different
one from that for the metallic atoms/ions. For the electrons
the transition
is one with increasing spatial disorder in the presence of the
electron-electron interaction. This means that 
for the electrons the disorder is of 
quenched type. This is so 
because of the adiabatic principle: On
their time scale the electrons experience a snapshot of the atomic arrangements.
These arrangements produce increasing spatial potential fluctuations
with decreasing density,
so that at a critical density
the Anderson-Mott transition takes place.

On the other hand, for the atoms/ions the transition scenario is not governed
by quenched disorder but by equilibrium thermodynamics.
The electronic degrees of freedom provide a density
dependent interaction.
Using the standard expression for the equation of states for
a simple liquid with a potential, which includes a
density-dependent screened
Coulomb term, we demonstrate that an unstable density interval
appears, which, in turn, produces the metal-nonmetal separation
in the liquid state. This mechanism of phase separation into
a metal-rich and metal-depleted liquid is very similar to that
suggested some time ago \cite{holzhey85,holzhey88,schirmacher15}
for solutions of metals in molten
salts \cite{bredig64,garbade88}.

In II. we introduce our model
and present the resulting equation of state. In section III.
we show isotherm calculations, which
are used to calculate a phase diagram. We conclude with discussing
achievements and shortcomings of our approach.

\section{Model}
\subsubsection{General formalism}
We start with the expression for the pressure
equation of states \cite{hansenmcdonald86,schirmacher15}
\be\label{eqstate1}
P(V,T)=\frac{k_BT}{V}\bigg(
1-\frac{1}{6}\frac{1}{Vk_BT}\int d^3\rr r\phi'(r)g(r)
\bigg)
\ee
where $g(r)$ is the radial distribution function, $T$ the temperature,
$k_B$ is Boltzmann's constant,
\mbox{$V=M/\rho_M$} the atomic volume, $\rho_M$ the mass density,
and $M$ the atomic mass.

We now assume that the interatomic potential is composed of 
three contributions:
\begin{itemize}
	\item[$(i)$]A hard-sphere contribution $\phi_{\rm hs}$;
	\item[$(ii)$]an attractive contribution $\phi_{\rm att}$.
	\item[$(iii)$]a screened-Coulomb contribution $\phi_{\rm sc}$;
\end{itemize}
We now lump the free-gas contribution to the pressure and the
hard-sphere potential together to a hard-sphere pressure
$P_{\rm hs}$ and write
\ba\label{eqstate2}
P(V,T)&=&P_{\rm hs}(V,T)-
\frac{1}{6V^2}\int d^3\rr r[\phi_{\rm att}'(r)+\phi_{\rm sc}'(r)]g(r)\nonumber\\
&=&P_{\rm hs}(V,T)+P_{\rm att}(V)+P_{\rm sc}(V)
\ea
For the hard-sphere pressure we use the 
Van-der-Waals repulsion term \footnote{See e.g. Ref. \cite{schirmacher15}
for the identification of Wan der Waals's $k_BT/(V-A)$ term with
the repulsive pressure.}
\be
P_{\rm hs}(V,T)=\frac{k_BT}{V-B}
\ee
with $B\approx d^3$, where $d$ is the 
distance of nearest approach or
effective hard-sphere
diameter. $B$ is also of the order of the atomic volume at melting.

Because the radial distribution function is strongly peaked near the
nearest-neighbour distance $d$,
and the potential contributions 
vanish for $r\gg d$ we may 
replace $\phi'(r)g(r)$ by a delta function and
approximately write

\be\label{delta0}
\frac{1}{V}\int d^3\rr r\phi_{\rm att;sc}'(r)g(r)
\approx Z(V)d\phi_{\rm att;sc}'(d)
\ee
with the coordination number
\be
Z(V)=\,\,\,\frac{1}{V}\!\!\!\!\!\!\int\limits_{|\rr|\leq r_Z}\!\!\!\!\!\! d^3\rr g(r)
\ee
where $r_Z$ is taken to be at the first minimum of $g(r)$.
It has been found experimetally \cite{winter91} that in some expanded
metals 
$Z$ increases linearly with density
\be
Z(V)=Z_0/V
\ee
with $Z_0\approx 8V_{\rm M,mp}$ for both expanded Rb and Cs,
where $1/V_{\rm M,mp}$ is the density at the melting
point. So we may write

\be\label{delta-app}
\int d^3\rr r\phi_{\rm att;sc}'(r)g(r)
\approx Z_0 d\phi_{\rm att;sc}'(d)
\ee
As generally the minimum of the attractive potential contribution is
located at $r_{\rm min}>d$, $\phi_{\rm att}'(d)<0$
and we get a van-der-Waals term
\be
P_{\rm att}(V)=-A\frac{1}{v^2}\, .
\ee
with
\be
A=\frac{Z_0}{6}r|\phi_{\rm att}(d)|\, .
\ee
Without the screening term
the equation of states $p_{\rm hs}(v)+p_{\rm att}(v)$ 
becomes the van-der-Waals equation of states, which gives
the
usual liquid-vapour transition scenario.

\subsubsection{Screening length and Metal-nonmetal transition}
We now turn to the main object of the present exercise, namely
the screened Coulomb potential. 

As indicated in the introduction we rely on the adiabatic principle,
from which follows that in the situation of an expanded metal the
(interacting) electrons experience a strongly spatially fluctuating
external potential due to the ion cores. These fluctuations are
``frozen'' on the time scale of the electrons ($\sim$ 1 fs). In
contrast to this the system of the metallic atoms/ions experience
effective pairwise
interaction potentials (see below),
which are
mediated by the electrons with their Mott-Anderson scenario.
If the electrons are in the metallic state, they are able to screen the
interionic Coulomb interaction.
The resulting effective interatomic interaction is
subject to equilibrium thermodynamics. So the electrons experience quenched
disorder, the atoms/ions experience annealed disorder. The 
resulting electronic
quenched-disorder Mott-Anderson transition is continuous, the thermodynamic one - as we shall see - is
discontinuous below the corresponding critical point.

The effective pairwise
interaction potential between the ions/atoms in a simple liquid
metal can be written
as the sum of a direct Coulomb repulsion and an indirect term, which
accounts for the screening
\cite{evans78,harrison82}:
\be\label{pot1}
\phi(r)=\frac{Q^2}{r}+\frac{1}{(2\pi)^3}
\int d\qq e^{i\qq\rr}\left|v_{ps}(\qq)\right|^2\chi_e(\qq)
\ee
Here $Q=N_ee$ is the ionic charge,
$e$ is the elementary charge, $N_e$ is the ionic relative charge
(number of electrons per ion or valence), 
$\chi(\qq)$ is the
electronic susceptibility and $v_{\rm ps}(\qq)$ the electron-ion
(pseudo) potential. $\chi_e(\qq)$ is the electronic susceptibility, which
in the Hartree approximation can be written as
\be\label{susc}
\chi_e(\qq)=\frac{q^2}{4\pi e^2}\frac{1-\epsilon(q)}{\epsilon(q)}
\ee
$\epsilon(q)$ is the Lindhard dielectric function of the free
electron gas \cite{ashcroft76}, which can be simplified using the
Thomas-Fermi approximation \cite{ashcroft76,harrison82}
\be
\epsilon(\qq)=1+\frac{\lambda_{TF}^2}{q^2}
\ee
with the Thomas-Fermi screening parameter (inverse squared screening length)
\be
\lambda_{TF}^2=4\pi e^2\mu_F
\ee
Here $\mu_F=4k_F/a_Be^2$ is the free-electron density of states at
the Fermi level, $a_B=\hbar^2me^2$ the Bohr radius, $m$ the electronic
mass and $k_F$ the Fermi wavenumber 
$k_F=\sqrt[3]{3\pi^2N_e/V}$.

Using this approximation for $\epsilon(\qq)$ and the empty-core
pseudopotential of \textcite{ashcroft66}, which is a Coulomb
potential $v_{ps}=-Q/r$ outside of the ionic radius $R_c$ and zero for $r<R_c$,
one obtains \cite{harrison82}
\be\label{pot}
\phi(r)=c(R_c)\frac{Q^2}{r}e^{-\lambda_{\rm TF}r}
\ee
where $c(R_c)$ is a prefactor related to $R_c$ \cite{harrison82}.

For the potential derivative we obtain
\be
r\frac{d}{dr}\phi(r)=-c(R_c)\frac{Q^2}{r}[1+\lambda(V)r]e^{-\lambda(V)r}
\ee

As mentioned in the introduction, the density of states can be considered
as the order parameter for the Mott-Anderson metal-nonmetal transition,
i.e. the transition of {\em interacting} electrons in the presence
of quenched disorder
\cite{belitz94,dobro97,byczuk05,dobro10,byczuk10}. In the typical-medium
DMFT treatment of the Mott-Anderson transition
the geometically averaged
local density of states $\mu$ vanishes linearly
with the control parameter, which is the width of the 
distribution of the fluctuating local potentials,
divided by the band width. As the latter is strongly density dependent,
we make the following ansatz for the local density of states

\be\label{critical1}
\mu(V)=\mu_F\,f\big(x(V)\big)
\ee
with the normalized density
$x(V)=\frac{\ds \VMNM}{\ds V}-1$ and
\be\label{critical2}
f(x)=x\theta(x)
\ee
where $\theta(x)$ is the step function and $\VMNM$ is the critical
atomic volume of the metal-nonmetal transition.

\begin{figure}
	\begin{center}
	\includegraphics[width=0.4\textwidth]{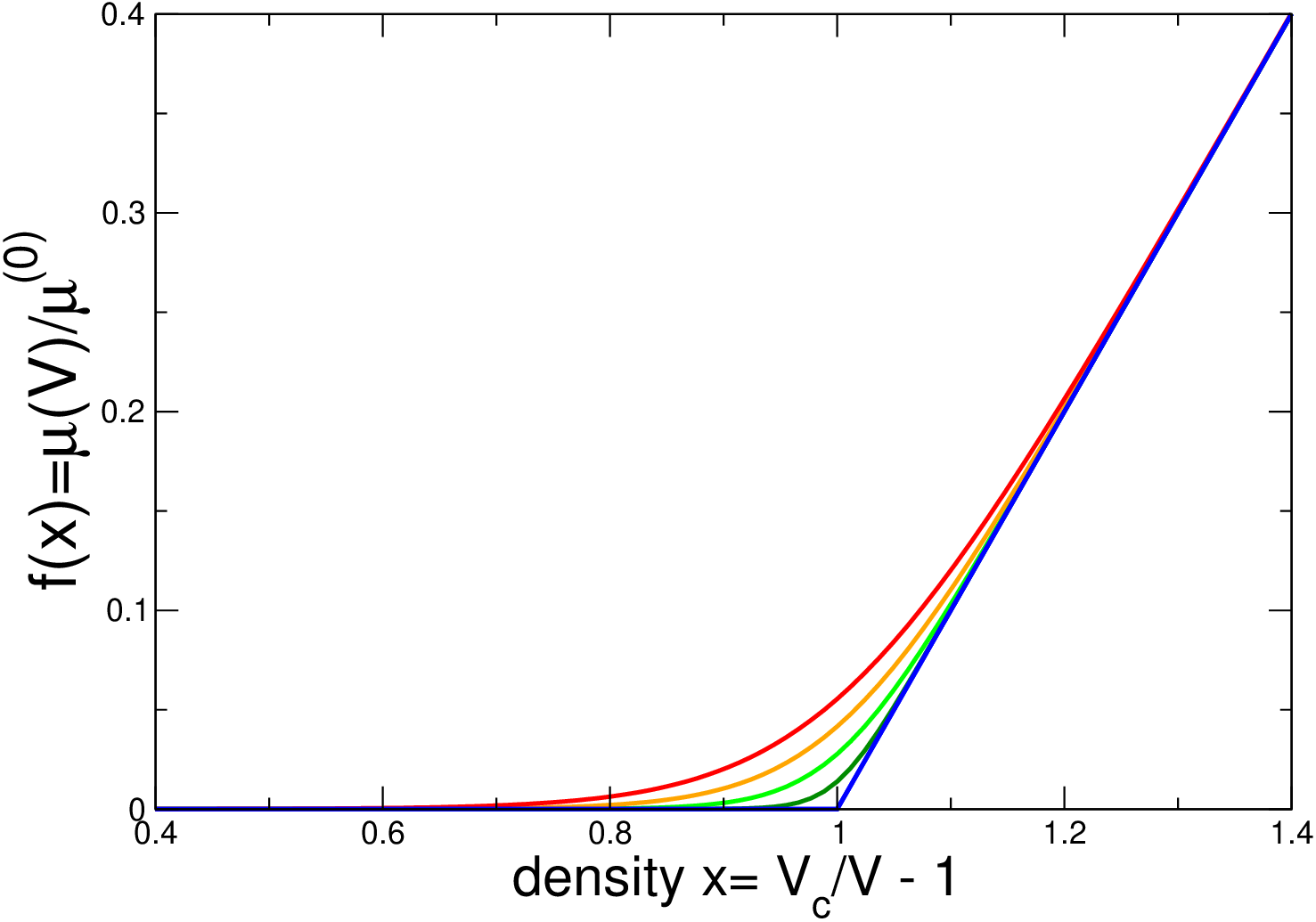}
\end{center}
	\caption{\small The function $\mu(v)/\mu_f^{(0)}$ of Eq. (\ref{critical2})
	for smoothing parameters $s$ = 0.0, 0.01, 0.02, 0.03, 0.04 
	from bottom (blue) to top (red).
	\label{smoothing}
	}
\end{figure}

So we have
\be\label{critical3}
\lambda(V)=\lambda_0(V)f^{1/2}[x(V)]
\ee
with $\lambda_0^2(V)=4\pi e^2\mu_F(V)$.
As the density, viz, volume dependence of $\lambda_0$ is considerably
weaker than the critical one we set $\lambda_0$ constant, i.e
$\lambda_0(V)=\lambda_0(V_{\rm mp})$. Finally
we may write
\be\label{eqstatescreen}
P_{\rm sc}(V)
=\frac{C}{V^2}\big[1+\lambda(V)d\big]e^{-\lambda(V)d}
\ee
with $C=c(R_c)Q^2Z_0/6d$. 

Collecting all the terms contributing to the pressure we obtain our
central result

\be\label{eqstate0}
P(V,T)=\frac{k_BT}{V-B}-\frac{A}{V^2}+\frac{C}{V^2}\big[1+\lambda(V)d\big]e^{-\lambda(V)d}\,\,.
\ee
Beyond the Anderson-Mott transition ($V>\VMNM$) we have
\be
P_{\rm sc}(V)
=\frac{C}{V^2}
\ee
so that in this limit we obtain an effective Van-der-Waals
equation of states with
\be
A_{\rm eff}=A-C
\ee
\begin{figure}
	\includegraphics[width=0.4\textwidth]{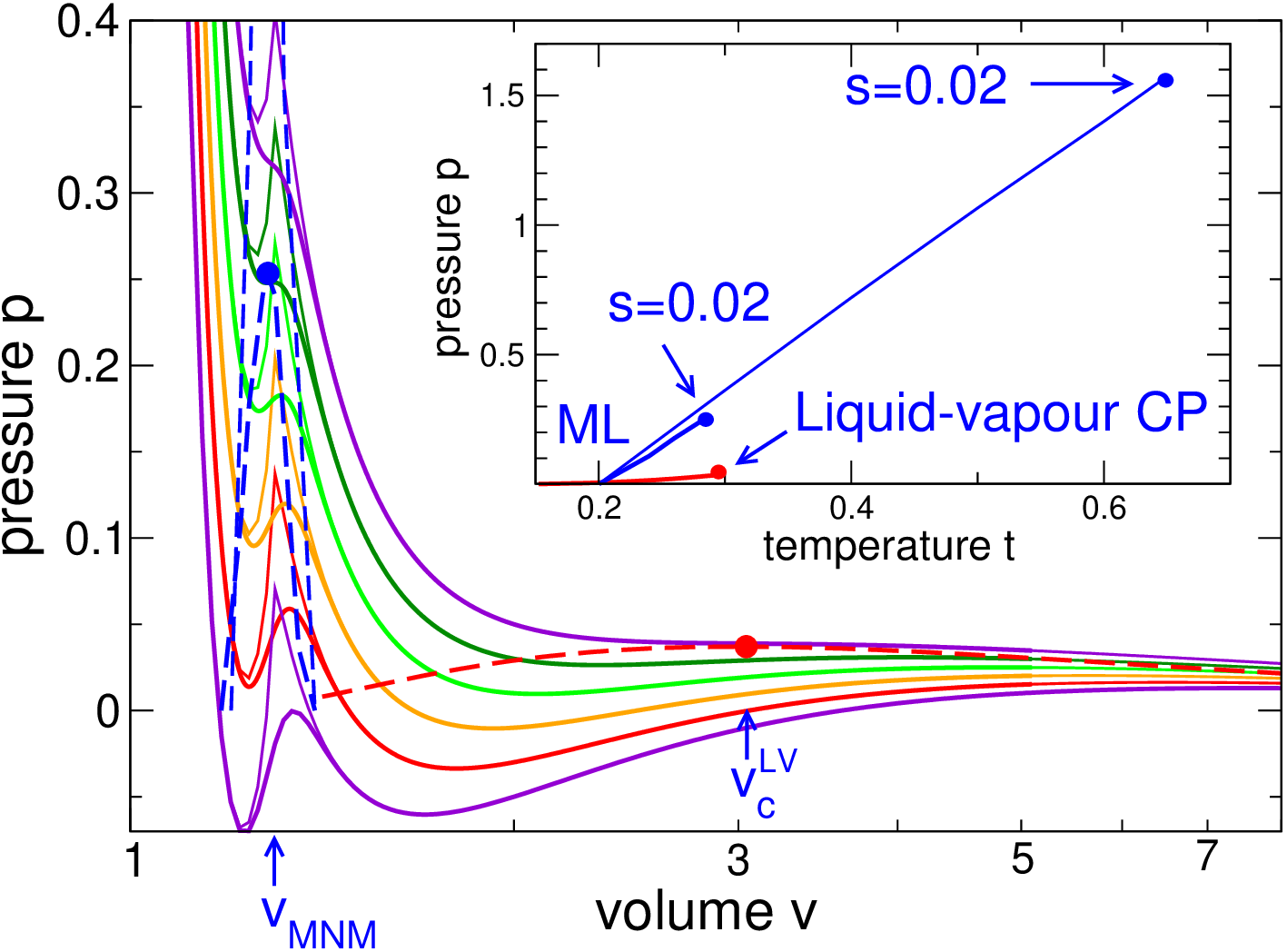}
	\caption{\normalsize $p-v$ isotherms according to our equation
	of states (\ref{eqstate}) for $zero$ smoothing parameter $s$
	(thin lines) and $s=0.02$ (thick lines) for the temperature range
	0.2 $\leq t \leq$ 0.3. 
	Further parameters used are $\vMNM$ = 1.3
	and $\lambda_0d=8$.
	The dashed lines are the equilibrium pressures calculated with the Maxwell and double-tangent construcion.\newline
	Inset: $p$ - $t$ phase diagram.\newline
	The dots indicate the critical points.
	\label{test2}
	}
\end{figure}

It should be noted that the equation of states (\ref{eqstate0}) interpolates
\cite{landaulifshitz80}
between that of a liquid metal (small $v$) and a free gas (large $v$).
\subsubsection{Inhomogeneities and smoothing of the metal-nonmetal transition}
Relation (\ref{critical3}) together with (\ref{critical2})
describes (at zero temperature) a rather
sharp transition between the metallic and nonmetallic state. Such
a transition is predicted by the generalized nonlinear sigma model,
which is based on weak disorder \cite{finkelstein83,belitz94,belitz95}.
At elevated temperature one may expect this transition to be somewhat
smoothed.

On the other hand, the alternative theory of the Mott-Anderson
transition, tailored for the case of strong correlations and strong
disorder 
\cite{dobro97,byczuk05,dobro10,byczuk10},
predicts a very broad distribution of local densities
of state $\mu_i$. A sharp transition is found for the
geometric average $\langle\mu\rangle_g=\exp\{\langle \ln\mu_i\rangle\}$,
wheras the arithmetic average $\langle\mu_i\rangle$ is non-critical.

We now phenomenologically introduce a {\it smoothing of the critical law}
\cite{vojta03}. We replace the function $f(x)$ in Eq. (\ref{critical1}),
which is the antiderivative of the step function
$\theta(x)$,
by the antiderivative 
$\widetilde f(x)$
of the complementary Fermi function $[1+e^{-x/s}]^{-1}$:
\be\label{critical4}
\mu(V)=\mu^{(0)}\widetilde f\big(x(V)\big)
\ee
with
\be
\widetilde f(x)=s\ln\big[1+e^{x/s}\big]
\ee
where $s$ is the smoothing parameter.
For $s\rightarrow$ 0 we recover (\ref{critical1}). 

In Fig \ref{smoothing} we show the influence of $s$ on the critical law.
As intended by construction the curves become increasingly smoother
with increasing $s$.
\section{Results}
We now use dimensionless units $v=V/B$, $t=k_BTB/A_{\rm eff}$ and
$p=PB^2/A_{\rm eff}$. In these units the equation of states takes
the form
\be\label{eqstate}
p(v,t)=\frac{t}{v-1}-\frac{1+c}{v^2}+\frac{c}{v^2}\big[1+\lambda(v)d\big]e^{-\lambda(v)d}\,\,.
\ee
with 
$c=C/A_{\rm eff}$

In these units the critical liquid-vapour quantities are given by
\be
t_c^{LV}=\frac{8}{27}\approx 0.3\qquad
\vMNM^{LV}=3\qquad
p_c^{LV}=\frac{1}{27}\approx 0.037
\ee

In Fig. \ref{test2} we show the isotherms predicted by the
equation of states (\ref{eqstate}) with the parameters indicated
in the caption. These isotherms show two unstable regimes:
at high atomic volume the usual liquid-vapour one, at low volume
the unstable regime
due to the density dependence of the screening, caused
by the (continuous!) metal-nonmetal transition.
As to be expected, the two-liquid instability occurs
in the vicinity of the critical Mott-Anderson volume $\vMNM$. 

\begin{figure}
	\includegraphics[width=0.4\textwidth]{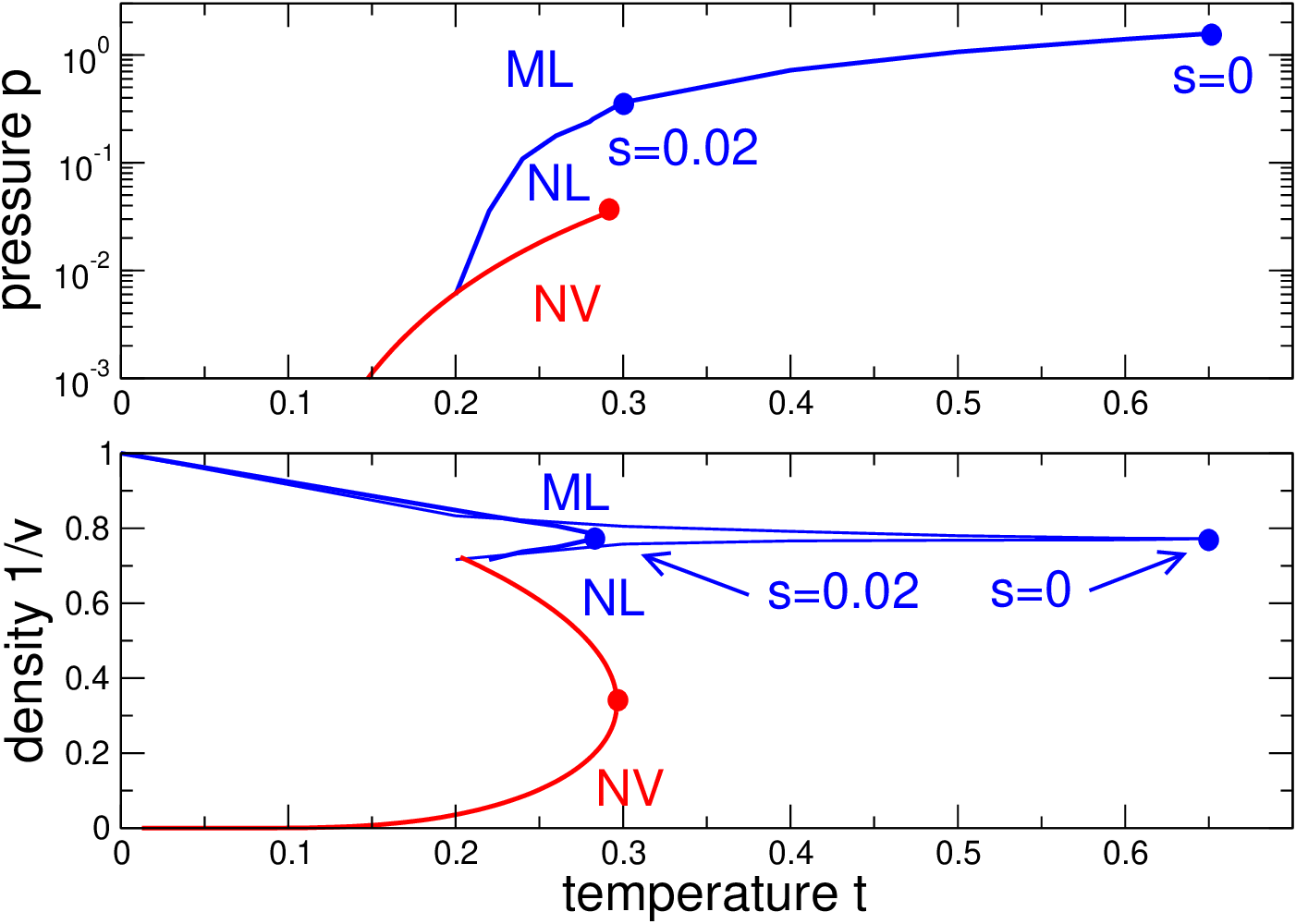}
	\caption{The pressure ($p$,$t$) and density ($v^{-1}$,$t$)
	phase diagrams corresponding to the isotherms of
	Fig. \ref{test2}. The dots indicate the critical points.
	ML = metallic liquid, NL = nonmetallic liquid, NV = 
	nonmetallic vapour.
	\label{rhophase}
	}
\end{figure}

We have
calculated the equilibrium volumes, pressures and temperatures using
both the Maxwell construction and the double-tangent method
\cite{huang01}. For the regime above $\vMNM$ we used Gibbs' parametric solution
of the Van-der-Waals coexistence problem \cite{lekner82}. Below
$\vMNM$ we implemented  a grapical double-tangent construction. In the immediate
vicinity of the critical point we used a numerical Maxwell construction, i.e.
equating the volumes above and below the coexistence pressure.
The resulting phase diagrams are shown in Fig. \ref{test2}
(dashed lines) and in \mbox{Fig. \ref{rhophase}.}

It is remarkable that the smoothing of the electronic
metal-nonmetal transition
results in a strong reduction of the critical endpoint
of the
corresponding atomic
transition. 

	Let us now consider the situation in expanded Hg and Rb.
We chose our parameter $\vMNM$, which is equal to $3\rho_c/\rho_{M\!N\!M}$
to be equal to 1.3, which corresponds to the value of $\rho_{M\!N\!M}/\rho_c$
= 2.3 in expanded Rb \cite{pilgrim18}. \textcite{pilgrim18} have evidence
by inelastic neutron scattering that in the density range around
$2.3 \rho_c$ a micro-emulsion of two liquids is present. 
Similar evidence has been presented earlier by \textcite{ruland09} for
expanded Hg in the range around $1.2\rho_c$ by analyzing published
small-angle scattering data \cite{inui06,inui07}. This would correspond to
$\vMNM$ = 2.5. It has been pointed out in Refs. \cite{ruland09,pilgrim18},
that the Coulomb interaction between the metallic micro-droplets, surrounded
by the second-phase nonmetallic material prevents a complete demixing
and establishes the micro emulsion.

\section{Discussion}
By combining the standard expression for the pressure equation of
state of a simple liquid with a hard-core repulsion, a short-range
attraction and a screened Coulomb potential reflecting the 
Mott-Anderson transition via a density-dependent screening length
we have constructed an equation of state, which gives rise
to a second coexistence line in the phase diagram.
As postulated by Zel'dovich and Landau \cite{landau43} we obtain
a separation into a metalic and nonmetallic liquid phase. Contrary
to their ideas and the ideas of Mott
\cite{mott49,mottdavis71,mott82,mott90}, we show that also
a {\it continuous} Mott-Anderson type metal-nonmetal transition
of the electrons gives rise to a {\it discontinuous} liquid-liquid
phase separation of the ions/atoms. We have introduced a phenomenologic
model for the density dependence of the screening length including
the possibility of a smoothed transition. We modeled the smoothing
or rounding of this transition \cite{dobro10,vojta03} by means of
the anti-derivative of the Fermi function featuring a smoothing
parameter. 

We find that the smoothing results in a reduction of the length of
the coexistence line.  Let us consider again the reasons for such a smoothing
to happen. First of all, the electronic transition does not take
place at zero temperature but at a temperature approaching the
Fermi temperature. Secondly, for $T=0$ we assumed the critical exponent
of the Mott-Anderson transition to be 1. If it would be larger than
one the curve would look like a rounded transition. Thirdly, 
as mentioned before, the local
density of electronic states in the Mott-Anderson scenario as given
by the dynamical mean-field theory 
\cite{dobro97,byczuk05,dobro10,byczuk10}
is known to exhibit strong spatial
fluctuations, so this as well will effectively lead to a smoothing
of the transition.
So a more detailed experimental investigation of the liquid-liquid
separation line will shed light on the details of the
mechanism of the Mott-Anderson transition.

	Finally we would like to discuss a point concerning the temperature
	dependence of our model. Our equation of states (\ref{eqstate0})
	has a linear temperature dependence like the van-der-Waals
	one. By elementary thermodynamic relations one can show that if the second
	temperature derivative (at constant volume) of the pressure is
	zero, such is the first volume derivative of the specific heat
	(at constant temperature). This implies that the specific heat
	does not depend on the volume. Of course, in a material, in which the
	electronic degrees of freedom play a dominant role, the 
	linear-temperature term of the specific heat should be present,
	which is proportional to the density of states at the
	Fermi level and, hence, should exhibit the same volume dependence as
	the screening parameter $\lambda(V)^2$. This is not included in our
	rather crude model. The model has mainly been introduced in order to 
	demonstrate, how a smooth metal-nonmetal transition can lead
	to a first-order phase transition in an expanded metal
	and a second coexistence line, as anticipated by Zeldovich and
	Landau \cite{landau43}. A more
	refined version should include a term quadratic in the temperature,
	which is then related to the specific heat. This term will
	be a correction of the order of $(k_BT/E_F)^2$, where $E_F$ is the
	Fermi energy. In a future publication we shall present a more refined version of our
	equation of states, in which the delta-function approximation
	of (\ref{delta0}) will not be made, and the $T^2$ term, related to
	the electronic specific heat, will be
	included.

\end{document}